\documentclass[a4paper,11pt]{article}
\pdfoutput=0 

\usepackage{jcappub} 

\usepackage[utf8]{inputenc}

\usepackage{amsmath,amssymb}
\usepackage[separate-uncertainty=true]{siunitx}

\usepackage{epsfig,graphicx}
\usepackage{graphicx}
\usepackage{subcaption}
\usepackage{wrapfig}
\usepackage{booktabs}
\usepackage{rotating}
\usepackage{cancel}
\usepackage{bm}
\usepackage{color}
\usepackage{comment}

\definecolor{green}{rgb}{0,0.5,0}
\definecolor{red}{rgb}{0.5,0,0}
\definecolor{blue}{rgb}{0,0,0.5}

\title{\boldmath First results from a hidden photon dark matter search in the meV sector using a plane-parabolic mirror system}

\author[a,b]{Stefan Knirck,}
\author[c]{Takayuki Yamazaki,}
\author[a]{Yoshiki Okesaku,}
\author[a]{\\Shoji Asai,}
\author[d]{Toshitaka Idehara,}
\author[c]{Toshiaki Inada}

\affiliation[a]{Department of Physics, Graduate School of Science, University of Tokyo, \\ 7-3-1 Hongo, Bunkyo-ku, Tokyo
	113-0033, Japan}
\affiliation[b]{Institut für theoretische Physik, Universität Heidelberg, \\ Philosophenweg 16, 69120 Heidelberg, Germany}
\affiliation[c]{International Center for
	Elementary Particle Physics, University of Tokyo, \\ 7-3-1 Hongo, Bunkyo-ku, Tokyo
	113-0033, Japan}
\affiliation[d]{Research Center for Development of Far-Infrared Region, University of Fukui, \\ 3-9-1
	Bunkyo, Fukui-shi, Fukui 910-8507, Japan}

\emailAdd{knirck@mpp.mpg.de}
\emailAdd{yamazaki@icepp.s.u-tokyo.ac.jp}
\emailAdd{tinada@icepp.s.u-tokyo.ac.jp}

\abstract{
We report on the first results from a new dish antenna search for hidden photon dark matter (HPDM) in the meV mass region.
A double mirror system composed of a plane and a parabolic mirror is designed to convert HPDMs into photons focused on a receiver. 
In this phase~1 experiment we obtain an upper limit on the photon-HP kinetic mixing $\chi \lesssim \num{e-8}$ for the mass range of  $0.67-\SI{0.92}{\milli\electronvolt}$ using conventional mm-wave technology with a room-temperature receiver and a small-sized mirror system.
}

\begin{document}
\maketitle
\flushbottom

\section{Introduction}
\label{sec:introduction}
For more than 40 years various astrophysical observations have suggested the existence of dark matter (DM)~\cite{Bertone:2004pz}.
Extensions of the Standard Model predict candidate particles classified as either weakly-interacting massive particles (WIMPs) with a typical mass of \si{\giga\electronvolt}--\si{\tera\electronvolt}, or weakly-interacting slim particles (WISPs) with \si{\micro\electronvolt}--\si{\electronvolt}~\cite{Arcadi:2017kky}.
Despite of intensive nuclear recoil experiments WIMPs have been escaping from detection so far,
while many ambitious projects have recently been proposed and proceed to search for WISPs, such as axions and hidden photons (HPs)~\cite{Graham:2015ouw,Irastorza:2018dyq,Jaeckel:2013eha, Bela:2016patras,TheMADMAXWorkingGroup:2016hpc}.
Especially, a wide area of the HP parameter space, defined by its mass $m_{\gamma^{\prime}}$ and the photon-HP kinetic mixing $\chi$, is still unexplored and allowed for the cold DM~\cite{Arias:2012az,Jaeckel:2013ija}.

Via the kinetic mixing HPDM can convert to ordinary photons with the energy directly corresponding to the HP rest mass within a narrow bandwidth $\sim \num{e-6} m_{\gamma^{\prime}}$, since cold DM moves at non-relativistic speeds $\sim \num{e-3} c$. 
This mixing is currently being investigated by using microwave resonant cavities~\cite{admx1, admx2, admx3, admx4, admx_hf, Graham:2015ouw,Irastorza:2018dyq}.
While the cavity experiments enhance the conversion power resonantly in a narrow frequency band,
the resonant frequency needs to be scanned to cover a broad mass region.
Recently, a new method using a metallic mirror has been proposed to obtain a broadband sensitivity over a full detector bandwidth~\cite{Horns:2012jf}.
In such a setup HPDM is converted at the surface of the mirror to ordinary photons. The electromagnetic boundary condition constrains the direction of emitted photons to perpendicular to the surface with a small angular spread $\sim\SI{e-3}{\radian}$, corresponding to the velocity dispersion of the incident HPs parallel to the conversion surface~\cite{Jaeckel:2013sqa}.
The emitted power is focused on a detector by using focusing optics, or using a curved mirror that works for both conversion and focusing.
A mirror with an area of $A_{\rm mirror}$ emits a HP power of
\begin{equation}
\label{eq:hppow}
	P = \alpha^2~\chi^2~\rho_{\rm HPDM}~A_{\rm mirror},
\end{equation}
with $\rho_{\rm HPDM}$ the local HPDM energy density and $\alpha$ a factor accounting for HP polarizations.
By assuming all polarizations are equally abundant\footnote{
	Even if HPDM is polarized in a particular direction, the earth's rotation would cause temporal variation of the direction observed in our laboratory frame.
	Therefore, the time averaging over more than a day results in a similar averaging over various polarizations.
}, we have $\alpha = \sqrt{2/3}$ and the generated photons have random polarizations.
Therefore, a detector sensitive to a power $P$ is sensitive to the kinetic mixing parameter $\chi$ as
\begin{equation}
\label{hpsens1}
\chi=4.5\times 10^{-9}\,\left(\frac{P}{10^{-13}\,{\rm W}}\right)^{\frac{1}{2}}
\left(\frac{1\,{\rm m^2}}{A_{\rm mirror}}\right)^{\frac{1}{2}}
\left(\frac{0.45\,{\rm GeV/cm^{3}}}{\rho_{\rm HPDM}}\right)^{\frac{1}{2}}
	\left(\frac{\sqrt{2/3}}{\alpha}\right),
\end{equation}
where the last two factors become unity under the Standard Halo Model\footnote{
Two different values have been used for the local DM density in literature.
We adopt 0.45 Gev/cm$^{3}$ in this work to be consistent with the previous works~\cite{admx1, admx2, admx3, admx4, admx_hf}.
The more commonly cited 0.3 GeV/cm$^{3}$ gives a reduction factor of 0.82 on the kinetic mixing sensitivity.
}~\cite{Horns:2012jf}.

Up to now, two experiments have probed masses around eV with this method primarily using photomultiplier tubes for the detection of optical photons~\cite{Suzuki:2015sza, Suzuki:2015vka, funk_ccd, Dobrich:2014kda, Dobrich:2015tpa,  Experiment:2017icw,Experiment:2017kmm}.
In Ref.~\cite{Suzuki:2015sza} the limit of $\chi \sim 6\times10^{-12}$ was obtained by using a 0.2 ${\rm m}^{2}$ dish antenna.
A search using a larger dish area of 13 ${\rm m}^{2}$ is also underway~\cite{Experiment:2017icw,Experiment:2017kmm}.
Besides, a radio frequency experiment is found to be in progress in the sub-meV region~\cite{Tajima:2016am}.

In this paper, we report a search for HPDM in the meV mass range corresponding to mm-wavelengths, and obtain the first limit in this region using the dish antenna method.
This mass region is particularly interesting, since a relatively large kinetic mixing, as high as $\chi\sim10^{-9}$, is still not excluded by cosmological and astrophysical observations~\cite{Arias:2012az}.
This work presents our phase~1 setup using heterodyne detection with a room-temperature receiver system and a small-sized dish antenna with an area of \SI{0.2}{\square\metre}.

\section{Experimental Setup}
\label{sec:exp-setup}

\begin{figure}[tb]	
\centering
\includegraphics[width=0.9\textwidth]{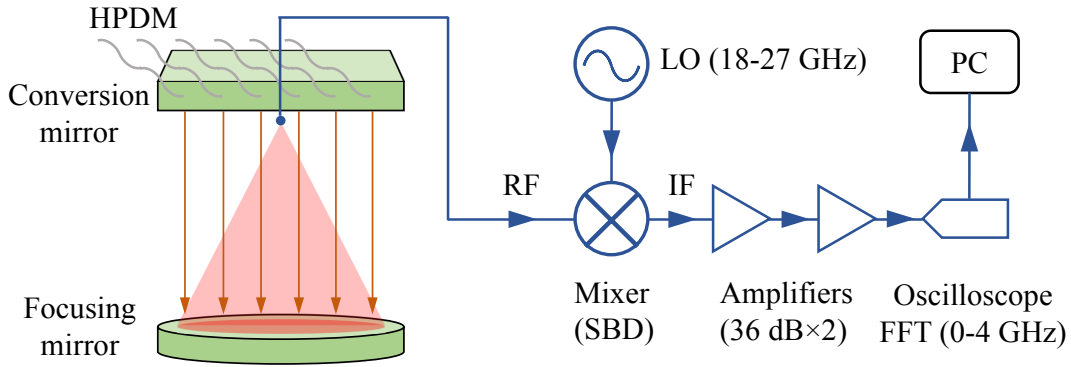}
\caption{
Schematic of the setup and the receiver chain.
Dark matter hidden photons are converted into ordinary photons at the surface of the plane aluminum mirror.
They are emitted in perpendicular direction to the surface and focused by the parabolic mirror.
A~horn antenna is placed at the focal point to receive the mm-wave power.
The RF frequency is down-converted by an even harmonic mixer (Schottky barrier diode) fed with a local oscillator signal at a frequency of 18--\SI{27}{\giga\hertz}.
The output IF power is amplified by two low-noise amplifiers, and the waveform is Fourier-transformed by an oscilloscope. 
The~power spectrum is stored after the linear average of $\num{e3}$ spectra.
\label{fig:schematic}}
\end{figure}

\begin{figure}[tb]	
\centering
\includegraphics[width=0.85\textwidth]{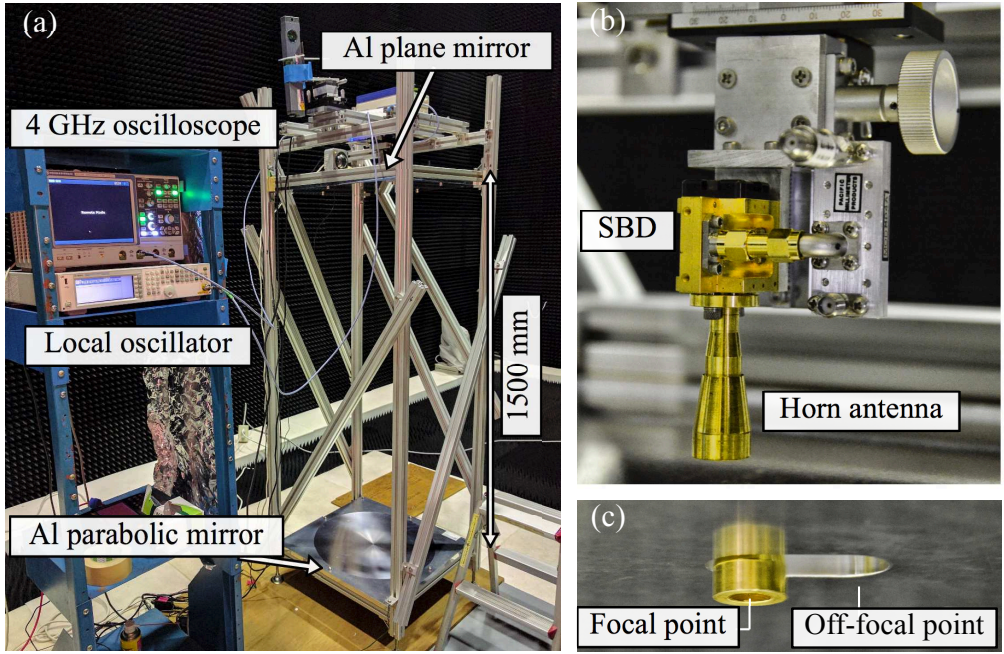}
\caption{
Pictures of the setup.
(a)~The measurement devices placed in a Faraday cage lined with microwave absorbers. 
During the data acquisition the mirror system is wrapped up with layers of crumpled aluminum foils and the other devices are moved to the next room or a distant place in the room to reduce radiation contamination.
(b)~The horn antenna attached to the Schottky barrier diode,
mounted on a multi-axis stage that adjusts the antenna position to the focal point.
(c)~Magnified view around the elongated hole at the center of the conversion mirror, where the horn antenna projects to receive the focal power.
\label{fig:photo}}
\end{figure}

Figure~\ref{fig:schematic} shows an overview of the experimental setup.
HPDM is converted to ordinary photons on a plane conversion surface and focused by a parabolic mirror.
Both mirrors are made of aluminum, which is commonly used for the mirror material of mm-waves due to its high reflectivity to these waves.
The conversion mirror has an area of \SI{0.6}{\metre}$~\times~$\SI{0.6}{\metre} and emits plane electromagnetic waves perpendicular to its surface.
The waves are focused by the parabolic mirror with a diameter of \SI{0.5}{\metre} and a focal length $f$ of \SI{1.5}{\metre}.
Placing the conversion mirror and the receiver at the same distance from the focusing mirror ($z=f$) minimizes the power loss due to the effective reduction of the conversion area.
This double mirror configuration 
has the advantage of being able to focus the signal power with a smaller directional spread compared to the setup using a single spherical mirror~\cite{Jaeckel:2015kea}.

The focused waves couple to a corrugated horn antenna shown in Fig.~\ref{fig:photo}.
The coupling to the antenna can be calculated by evaluating the overlap integral between the focused beam and the antenna beam \cite{antenna_coupling}. This yields a power coupling
\begin{equation}
C = \left( \frac{2 \, \omega_{0} \, \omega_{b} }{
{\omega_{0}}^2 + {\omega_{b}}^2}
 \right) ^{2},
\label{eq:coupling}
\end{equation}
where $\omega_{0}=\SI{3.0}{\milli\metre}$ is the beam size at the horn aperture giving the best coupling.
The focal beam size $\omega_{b}$ is determined by the successive diffraction at the two mirrors and becomes about $3.3 \lambda = 4.5-\SI{6.2}{\milli\metre}$ in the SBD input range (defined later)~\cite{stefan_ms}.
From Eq.~\eqref{eq:coupling}, this corresponds to $C=0.62$--$0.85$.

The antenna is attached to the Schottky barrier diode (SBD, Virginia Diode WR5.1EHM).
The SBD works as an even harmonic mixer and down-converts an RF frequency $f_{\rm RF}$, related to the HP mass with $m_{\gamma^{\prime}} [{\rm \mu eV}] \simeq 4.1\times (f_{\rm RF}[{\rm GHz}])$, into an IF frequency $f_{\rm IF}$ as
\begin{equation}
f_{\rm IF} =  | f_{\rm RF} - N f_{\rm LO} |,
\label{eq:sbd}
\end{equation}
with $N(=8)$ the even harmonic number we used and $f_{\rm LO}$ a local oscillator (LO) frequency fed by a signal generator (Agilent N5183A) with high frequency stability.
The modulus in Eq.~\eqref{eq:sbd} states that the SBD operates as a double sideband receiver, that effectively doubles the sensitive mass region around the center frequency $N f_{\rm LO}$.
The conversion loss of the SBD was measured before the HP search at 30 frequencies between \SI{160}{\giga\hertz} and \SI{220}{\giga\hertz} (``SBD input range''), by using a backward wave oscillator (Microtech, QS1-260-OV-24) as a variable-frequency mm-wave source.
It was found to vary between \SI{30}{\decibel} and \SI{50}{\decibel} in the range\footnote{
In this pre-measurement, we kept the LO power constant, which is the same condition as the HPDM search.
Optimizing the LO power to each $N$ and $f_{\mathrm{RF}}$ would give better (smaller) conversion loss~\cite{virginia_diode}.
}.
The SBD couples to a linear polarization and thus effectively detects a half of the incident HP power on average, since the photons emitted from the conversion surface have random polarizations.

The output power from the SBD is increased by two amplifiers (R\&K LA130-0S), whose gain curves were measured using a vector network analyzer (Anritsu, MS2024B). After the amplifiers, an oscilloscope (Rhode\&Schwarz RTO1044) records the FFT spectrum up to \SI{4}{\giga\hertz} with a resolution of \SI{76}{\kilo\hertz}.
Each spectral data is stored after the linear average of $\num{e3}$ spectra, which takes \SI{80}{\second}.

\begin{table}[!t]
	\centering
	\begin{tabular}{cccccc}
		\hline
		ID & Date & $f_{\rm LO}$ & $N_{f_{\rm LO}}$ & Background & $N_{\rm A}$ \\ \hline \hline
		1 & 11.04--11.28 & $\SI{23.17}{\giga\hertz}$ & 1 & Off-focal & \num{11831} \\
		2 & 01.16--01.26 & $(18, 19, ..., 26, 27)\, \si{\giga\hertz}$ & 10 & LO detune & \num{577} \\
		3 & 01.26--01.31 & $(18, 19, ..., 26, 27)\, \si{\giga\hertz}$ & 10 & LO detune & \num{294} \\
		4 & 01.31--02.03 & $(20, 21, ..., 26, 27)\, \si{\giga\hertz}$ & 10 & LO detune & \num177{} \\
		5 & 02.03--02.13 & $(20, 20.25, ..., 27, 27.25)\, \si{\giga\hertz}$ & 16 & LO detune & \num{354} \\
		6 & 02.17--03.31 & $(20, 20.25, ..., 27, 27.25)\, \si{\giga\hertz}$ & 16 & Off-focal & \num{1338} \\
		\hline
	\end{tabular}
	\caption{
		Summary of the data sets.
		The first data were taken in Nov.~2016, while the rest from Jan.~to Mar.~2017 with various measurement conditions, changing the frequency of the local oscillator ($f_{\rm LO}$) and background measurement mode.
		$N_{f_{\rm LO}}$ is the number of LO frequencies scanned with an interval of 1 GHz, while in the fifth and sixth data sets, the two intervals of \SI{0.25}{\giga\hertz} and \SI{0.75}{\giga\hertz} were repeated alternately to cover the SBD input range seamlessly.
		One data set contains $N_{\rm A}$ spectra for each $f_{\rm LO}$.
		Each spectral data stored is the linear average of $\num{e3}$ spectra.
		Thus, a data set contains a total of $N_{f_{\rm LO}} N_{\rm A}$ averaged spectra for both signal and background measurements.
		\label{tab:runs}}
\end{table}

To cut radio contamination from external sources, the HP search was carried out in a dedicated radio-shielding room (``Faraday cage'') at the Research Center for Development of Far-Infrared Region, University of Fukui (FIR-UF) at~N~\ang{36;4;37}~E~\ang{136;12;42}. It attenuates the power of external radio sources by \SI{60}{\decibel}, while the polyethylene foams lining the room provide an additional absorption of \SI{40}{\decibel}. 
Radiation from our measurement devices is reduced by leaving the minimal electronics in the room while the rest is placed in the next room that has a connection to the Faraday cage through a port duct.
In addition, the mirror system is wrapped up with layers of crumpled aluminum foils to further cut background radiation.
Even in this condition, a small amount of radiation contaminates and appears in the measured spectra.
These spurious peaks are discriminated by two types of background measurements, where the horn antenna is moved to an off-focal position by a motorized stage, or the LO frequency is detuned by $\Delta f_{\rm LO}=+\SI{1}{\mega\hertz}$ with the antenna at the focal position.
If spurious peaks are real signal, they would disappear in the off-focal measurement, or shift by $N\Delta f_{\rm LO}$ in the detune measurement.
Furthermore, the combination of the two background measurements enables to identify whether the radiation is mixed in the IF chain or more upstream paths.
To cancel short-term temperature dependencies of these peaks, signal and background acquisitions were repeated alternately for each 80 s.
These changes are remotely controlled by a PC placed in the next room without disrupting the measurement condition by entering the Faraday cage.

\section{Data Analysis}
\label{sec:dataanalysis}
\begin{figure}[!t]	
	\centering
	\includegraphics[width=0.8\textwidth]{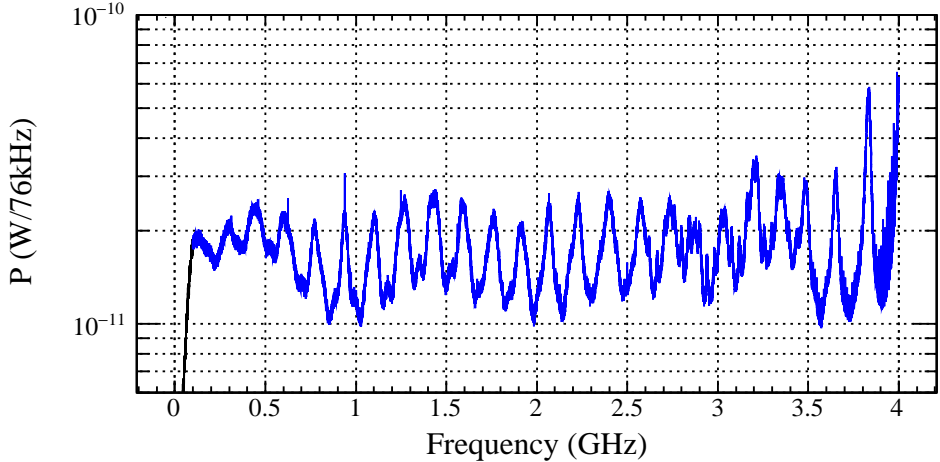}
	\caption{
		Power spectrum of a single acquisition ($10^{3}$ averages) obtained as the first spectrum of the last data set by the signal measurement with a local oscillator frequency of \SI{27}{\giga\hertz}, normalized with the receiver system gain.
		The steep structure in the low frequency region represents the amplifier cut-off.
		Thus, the frequency region of $0.1-\SI{4.0}{\giga\hertz}$ shown in blue is used for the analysis.
		The baseline oscillation becomes larger at high frequencies.
		The apparent peaks, for example at \SI{0.6}{\giga\hertz}, \SI{0.9}{\giga\hertz}, and around \SI{3.9}{\giga\hertz}, disappear after locally subtracting the spectrum of the corresponding background measurement in the region around these peaks.
	}
	\label{fig:raw}
\end{figure}

Table~\ref{tab:runs} shows a summary of the data sets.
Signal peaks are searched in the frequency domain of oscilloscope spectra.
A typical power spectrum of a single acquisition is shown in Fig.~\ref{fig:raw}, that has been normalized with the receiver system gain described in Sec.~\ref{sec:exp-setup}.
The overall spectral height represents the noise level of the receiver system.
The broad structure that oscillates with a frequency of $100-\SI{200}{\mega\hertz}$ is most likely to arise from power resonances, such as in RF cables.
The expected mm-wave HPDM signal has a sharp Maxwellian distribution over a few bins due to the DM velocity dispersion and is 
added on top of white noise on the baseline oscillation.
To extract the noise component, the spectrum is fitted to a quadratic function for 40-bin intervals and subtracted by the obtained baseline function~\cite{admx1}.
The resulting spectrum contains white noise and peaks having large positive power, as shown in Fig.~\ref{fig:raw}. 
These peaks are spurious and originate from residual radio contamination, because they also appear at the same IF frequency even in the LO-detuned background spectrum.
These spurious peaks are eliminated by subtracting the spectrum of the corresponding background measurement. 
The subtractions are operated locally around the spurious peaks to keep the noise level smaller by a factor of $\sqrt{2}$ for most of the white spectrum, compared to subtraction of the entire background spectrum.

\begin{figure}[t]
	\begin{minipage}{0.5\hsize}
		\centering
		\includegraphics[clip,width=0.95\textwidth]{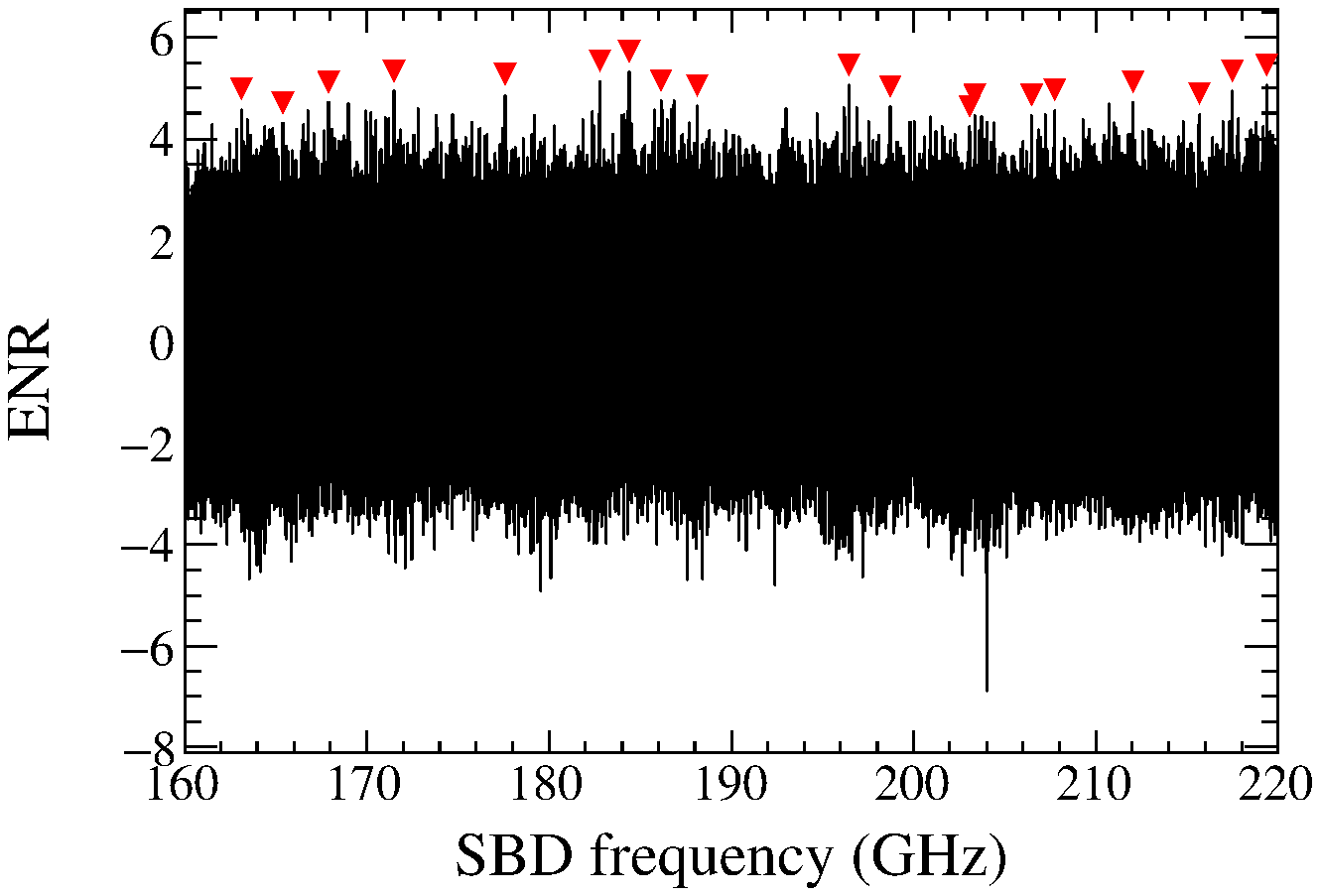}
	\end{minipage}
	\begin{minipage}{0.5\hsize}
		\centering
		\includegraphics[clip,width=0.95\textwidth]{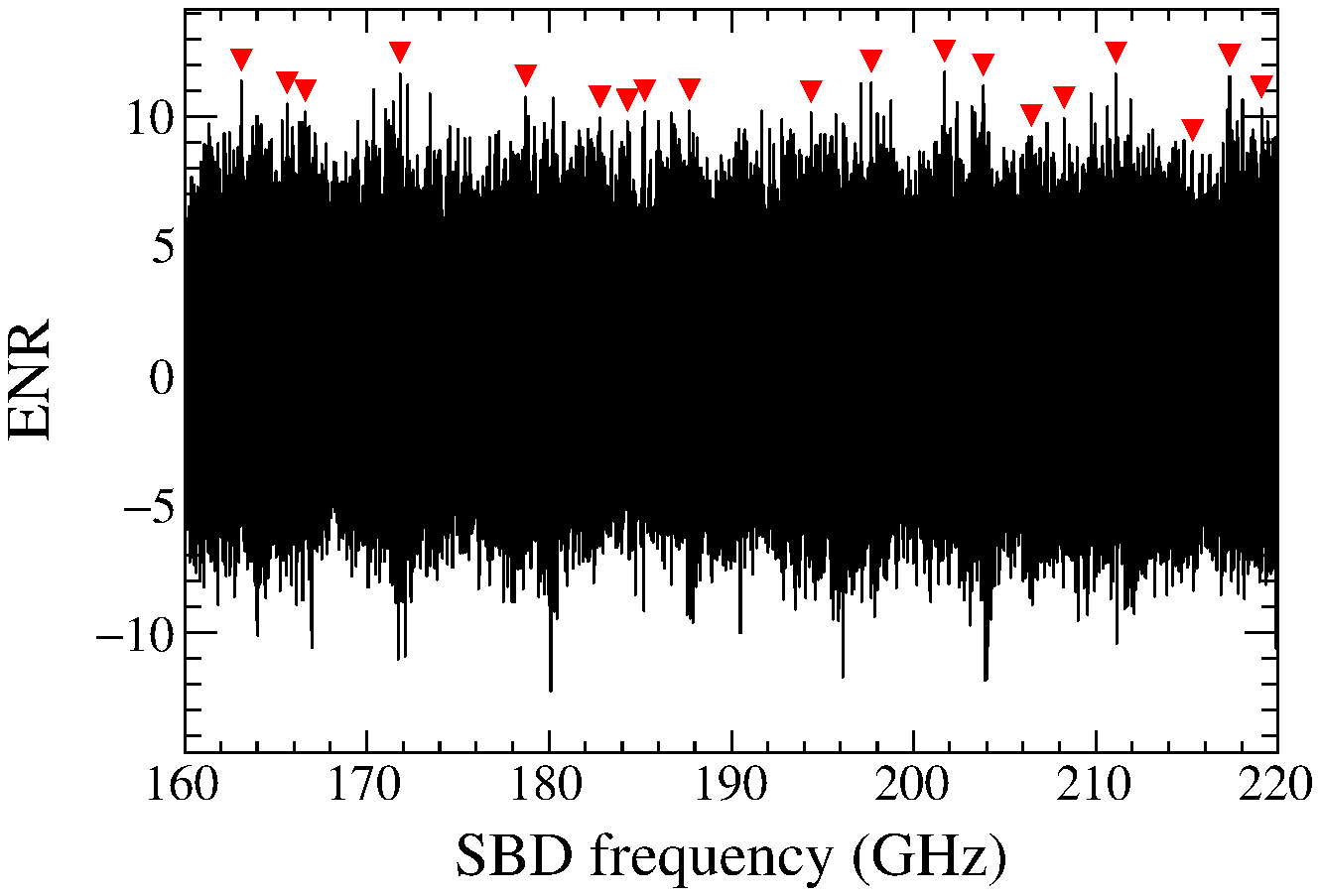}
	\end{minipage}
	\caption{
		Excess over noise ratio (ENR) of the power spectrum for the single-bin (left) and four-bin (right) analyses.
		The red triangles show positions of the largest ENR in sub-regions, decided by the pre-measured frequency dependency of the receiver system gain.
		\label{fig:enr}}
\end{figure}

One data set contains $N_{\rm A}$ acquisitions for each $f_{\rm LO}$, and they are combined with linear weighting according to the local power level of the baseline spectrum~\cite{admx2}.
The combined spectra are mapped to $f_{\rm RF}$ with Eq.~\eqref{eq:sbd}.
To cover the SBD input range without any sensitivity gap, $f_{\rm LO}$ is changed alternately between the two intervals of \SI{0.25}{\giga\hertz} and \SI{0.75}{\giga\hertz} in the last two data sets in Tab.~\ref{tab:runs}.
Finally, the mapped spectra of all data sets are combined to a single spectrum shown in Fig.~\ref{fig:enr}~(left),
where the power excess has been normalized to the local rms noise power, giving the excess over noise ratio (ENR).

For gravitationally thermalized (``virialized'') HPDMs, the Maxwellian signal spreads to about four bins, that contain 84--93\% of the total power depending on the HP mass frequency in the SBD input range.
To sum up the signal power, we apply a co-adding scheme, that simply takes the sum over the four adjacent bins to each frequency bin~\cite{admx3}.
This approach is robust since it does not assume any specific line shape on the signal discrimination, and is thus valid even if the actual local DM velocity distribution is not Maxwellian or contains some substructure~\cite{OHare:2017yze}.
Figure~\ref{fig:enr} (right) shows the co-added ENR spectrum.
According to the frequency dependence of the SBD conversion loss measured at many points (Sec.~\ref{sec:exp-setup}),
we divide the whole SBD input range into 19 sub-regions, where the conversion loss is regarded as constant.
Generally, candidate peaks above a certain ENR threshold are investigated by additional measurements that test whether they are truly statistical fluctuations.
While the rescanning measurement contributes to reject noise fluctuations~\cite{admx1, admx2, admx3},
we get around this step by taking the largest ENR as the threshold in each sub-region.
They are commonly distributed in the range of 9--12 ENRs, beyond which no significant peaks are observed. Since we expect one or more peaks above the highest observed ENR within the whole spectrum with a probability of $\approx 14\%$, non of the observed peaks is significant enough to claim a signal.
In order to obtain a limit on the observed HPDM power, the power level that gives ENRs beyond the threshold with a probability of 90\% was estimated by a Monte Carlo (MC) simulation, since neighboring bin contents are correlated.
The MC spectra are produced in a data driven manner, where white Gaussian noise and a Maxwellian signal are superimposed to the baseline spectrum of real data~\cite{admx1, admx2, admx3}.
Changing the signal power for 4000 masses randomly selected in the SBD input range, we applied the same analysis to the MC data and obtain a power limit in each sub-region, which is typically $P\sim 10^{-13}$ W.
If HPDM has a very small velocity dispersion in contrast to the above virialized scenario, it would deposit all power into a single bin.
In this ``unvirialized'' case, the power limits are directly calculated from the single-bin thresholds in Fig.~\ref{fig:enr}~(left) by using conventional Gaussian statistics~\cite{admx4}.

%
Furthermore, our analysis accounts for systematic uncertainties 
arising from
(i)~uncertainties on the local HPDM momentum distribution, 
(ii)~geometrical imperfections of the mirror system,
(iii)~uncertainties on stability and gain of the detection system electronics.
(i)~As described above, our result is unaffected if the local DM velocity distribution is not Maxwellian or contains substructure. However, the motion of the solar system in our galaxy induces the so-called ``DM wind'' at $\sim \SI{220}{\kilo\meter\per\second}$ in the rest frame of the earth. The direction of the wind observed in the laboratory frame changes in time due to the revolution and the rotation of the earth.
This effect shifts the overall direction of perpendicularly emitted photons by at most $\num{e-3}$~rad ($= \theta_{\rm{w}}$)~\cite{Jaeckel:2015kea}. It leads to a displacement of the focal spot by $f \theta_{\rm w}$ with $f$ the focal length ($=\SI{1.5}{\metre}$). This gives a smaller coupling to the horn antenna and results in a power reduction by a factor of up to~$0.85$.
(ii)~Non-parallel mirrors cause another displacement of the focal point by $f \theta_{\rm t}$ with $\theta_{\rm t}$ the tilt angle between the mirrors.
To check it, we developed a surveying method utilizing multiple reflections of a \SI{650}{\nano\metre} laser beam, where the displacement is augmented as the beam bounces between the mirrors~\cite{stefan_ms}.
We repeated this method during the run and confirmed that $\theta_{\mathrm{t}} < \SI{2(2)e-4}{\radian}$ throughout the data acquisition, which is significantly smaller compared to $\theta_{\rm{w}}$.
The focal power would also be reduced by the surface roughness of the two mirrors, causing phase errors of the emitted and reflected mm-waves.
The phase error becomes important especially to the parabolic mirror since it requires elaborate surface processing.
The surface roughness was measured to be less than \SI{30}{\micro\metre} by the manufacturer and is thus much smaller than the mm-wavelengths we receive.
The corresponding phase error is estimated to reduce the power by a factor of $0.97$~\cite{phase_error1, phase_error2}.
(iii)~The frequency stability of the detection electronics is limited by the signal generator. It has a temperature dependence of $\pm 1$~ppm over $0-\SI{55}{\degreeCelsius}$. 
The monitored room temperature shows seasonal changes in the range of $\pm \SI{3}{\degreeCelsius}$ throughout the search, that corresponds to the fluctuation in $f_{\mathrm{IF}}$ of about $\pm \SI{10}{\kilo\hertz}$.
It causes broadening of the Maxwellian power spectrum, giving at most 1.5\% of the power spilling out of co-added four bins ($4 \times \SI{76}{\kilo\hertz}$).
The largest systematic uncertainty in the receiver system is found in the total gain after the SBD and the two amplifiers.
As described in Sec.~\ref{sec:exp-setup}, we carried out these studies before the HPDM search and its uncertainty is already included in the analysis by conservatively using the minimum gain in each sub-region.

\label{sec:conclusion}
\begin{figure}[!t]
	\begin{minipage}{0.5\hsize}
		\centering
		\includegraphics[clip,width=0.95\textwidth]{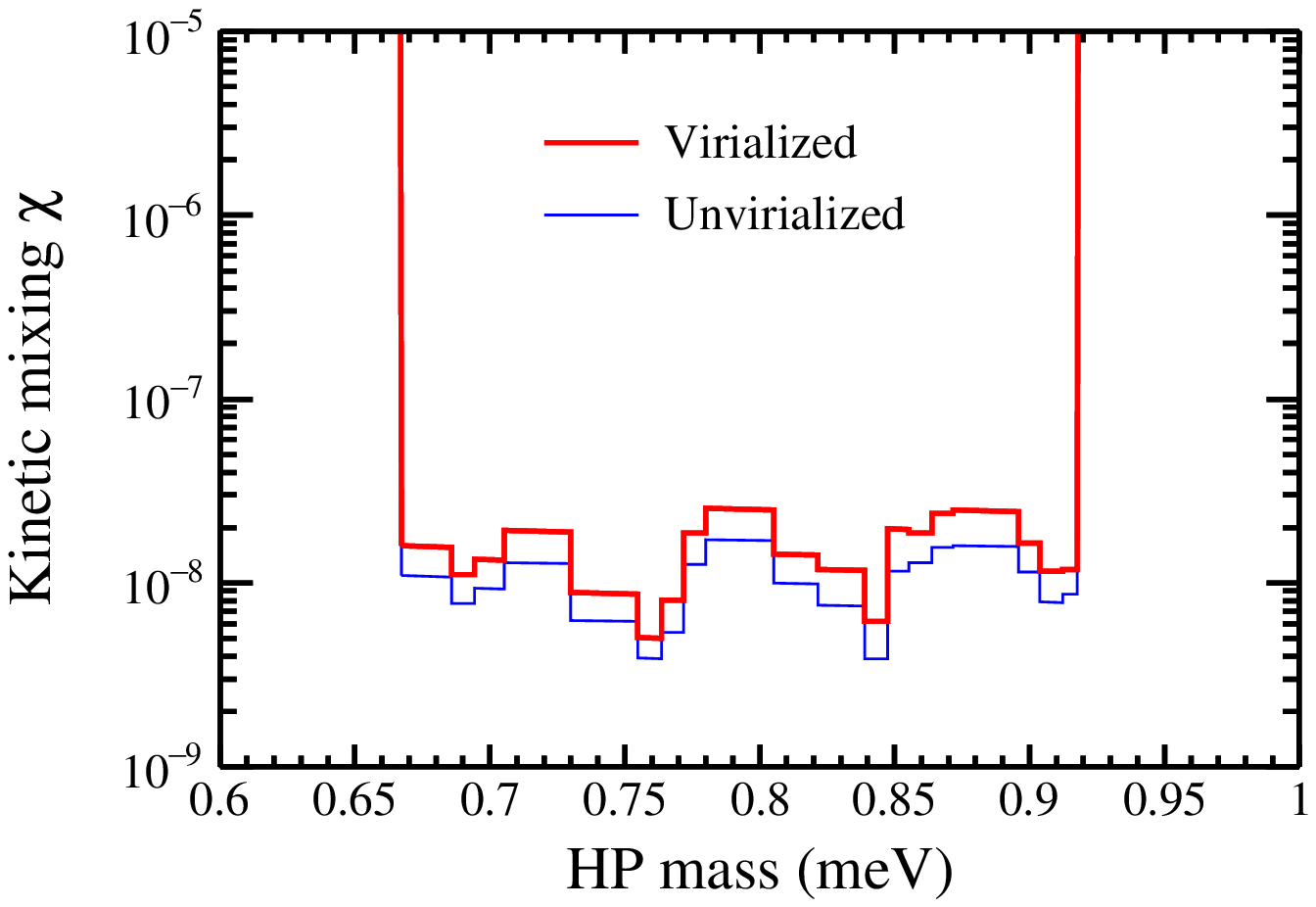}
	\end{minipage}
	\begin{minipage}{0.5\hsize}
		\centering
		\includegraphics[clip,width=0.95\textwidth]{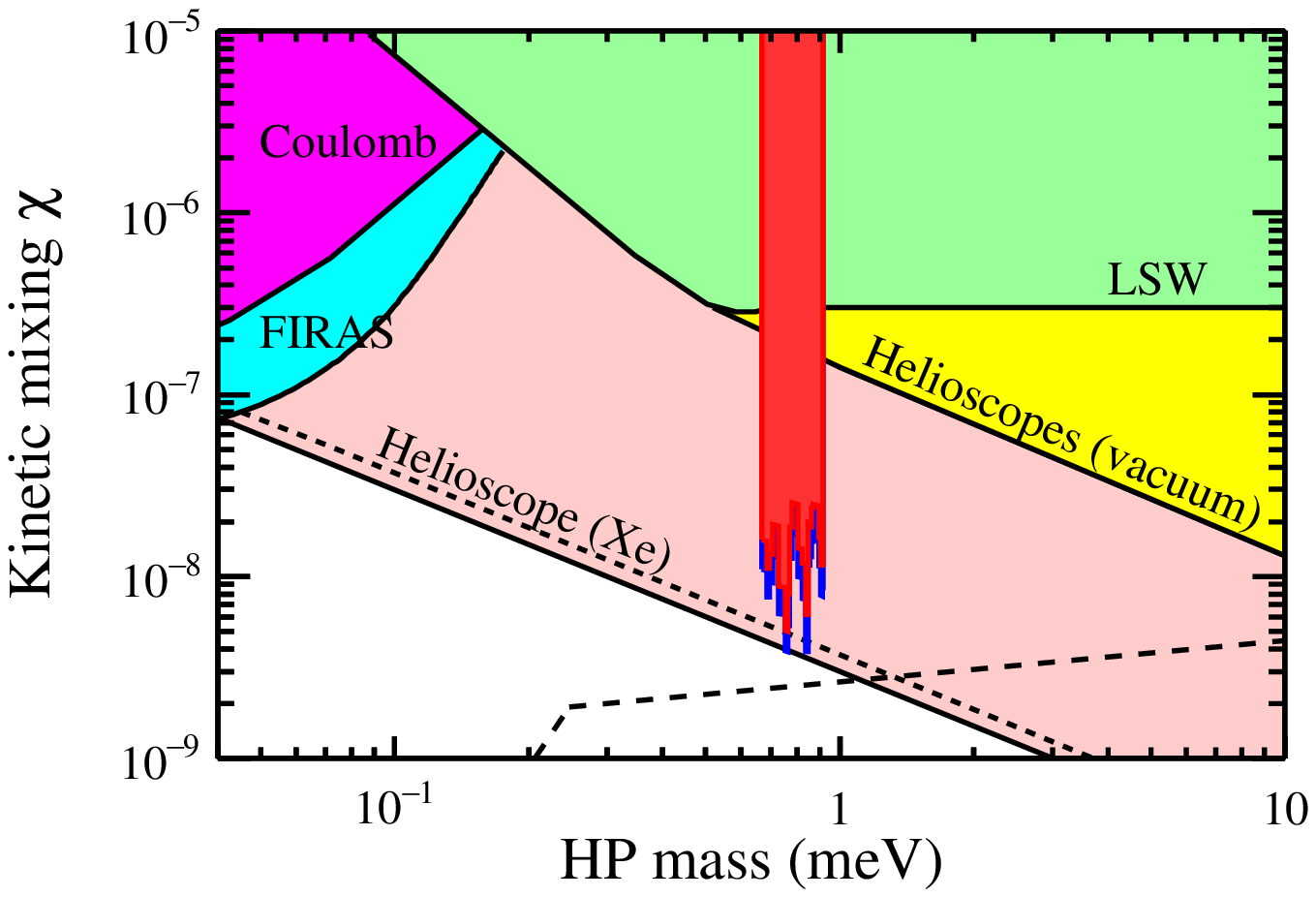}
	\end{minipage}
	\caption{
		Limits on the photon--HP kinetic mixing.
		(Left) Limits obtained in this work at a 90\% confidence level. The thick and thin lines are virialized four-bin analysis and unvirialized single-bin analysis, respectively.
		(Right) Comparison of the result with other limits;
		"LSW" light-shining-through-a-wall experiments using a laser~\cite{alps},
		"Coulomb" a test of deviation from Coulomb's law~\cite{coulomb1, coulomb2},
		"FIRAS" a test of spectral distortion of the cosmic microwave background~\cite{firas},
		"Helioscopes~(vacuum)" solar HP searches in vacuum~\cite{Redondo:2008aa,sumico,Schwarz:2015lqa},
		"Helioscope~(Xe)" a solar HP search for xenon photoionization~\cite{xenon}.
		Dotted line is a constraint from the solar energy loss accounting for longitudinally polarized HPs~\cite{solar_l1, solar_l2}.
		Dashed line is a cosmological constraint from the effective number of neutrino species~\cite{Arias:2012az}.
		\label{fig:limits}}
\end{figure}  

Applying Eq.~\eqref{hpsens1}, we obtain the limits on $\chi$ shown in Fig.~\ref{fig:limits}~(left).
The stepwise structure mostly represents the difference of conversion loss among the sub-regions, and partially the variation of power thresholds and the amount of statistics among the data sets.
We compare the result with other limits in Fig.~\ref{fig:limits}~(right).
The filled areas show the excluded regions from direct measurements.
Our search provides new limits, that are almost compatible to the astrophysical constraints from the solar energy loss argument and from the solar HP search in xenon experiments~\cite{solar_l1, solar_l2, xenon}.
While their estimation of the HP flux on the earth inevitably depends on a solar model, the limit obtained here postulates only the local DM energy density on the HP flux.
This work also shows an example of the double mirror configuration in HPDM searches, that can be applied to other bands of radio frequencies.

\section{Conclusion and Prospects}

This work demonstrates the dish antenna method using mm-wave technology applied to a HPDM search and obtains the first haloscope limit in the meV mass region.
We exclude the kinetic mixing $\chi \gtrsim \num{e-8}$ at a 90\% confidence level for the mass range of $0.67-\SI{0.92}{\milli\electronvolt}$.

One way to improve the sensitivity is to lower the power threshold by additional rescanning measurements omitted in this work.
$f_{\rm LO}$ will be tuned close to the frequency of each candidate peak to intensively take data with a low-noise narrow-band receiver.
Supposing that the ENR threshold is reduced by a factor of $R$ in a reasonable run time, it lowers the limit on $\chi$ by $R^{1/2}$.
Moreover, we note that decreasing 
the conversion loss by using other harmonics $N$ or by optimizing the LO power may allow for significant improvements in future.
To improve the sensitivity more drastically, one can use a cryogenic receiver applying a superconducting tunnel junction, and a low-temperature preamplifier.
This would reduce the noise compared to the current room-temperature system by many orders of magnitude.
Furthermore, coating dielectric layers on the mirror surface has been pointed out to boost the emitted power~\cite{Jaeckel:2013eha,TheMADMAXWorkingGroup:2016hpc}.
This interesting and promising scheme is currently being studied in a proof-of-concept phase as in Refs.~\cite{Bela:2016patras,TheMADMAXWorkingGroup:2016hpc}.

\appendix










\acknowledgments
The authors would like to thank Joerg Jaeckel, Tomoki Horie, Jun'ya Suzuki, Yoshizumi Inoue and Makoto Minowa for interesting discussions and helpful comments.
This work was funded in part by JSPS KAKENHI (Grant No. JP17K14269).
We would also like to thank KEK Computing Research Center for providing infrastructure for the Monte Carlo simulation.



\end{document}